\documentclass[10pt,twocolumn,showpacs,showkeys,preprintnumbers,amssymb,aps,superscriptaddress,pre]{revtex4-2}

\usepackage[colorlinks={true}]{hyperref}
\hypersetup{colorlinks=true,linkcolor=red,citecolor=magenta,urlcolor=blue}
\usepackage{color,array}
\usepackage{amsmath, amssymb}
\usepackage{graphicx}
\usepackage{bm}
\usepackage{wrapfig}
\usepackage{subfigure}
\usepackage{epsfig}
\usepackage{epstopdf}
\usepackage{relsize}
\usepackage{dblfloatfix}
\usepackage{orcidlink}
\usepackage{pstricks}
\usepackage{enumitem}
\usepackage{fixltx2e}
\usepackage{float}
\usepackage{placeins}
\usepackage{afterpage}
\usepackage{xcolor} 

\begin{document}


\title{Molecular Nanomagnet $\text{Cu}^\text{II}\text{Ni}^\text{II}\text{Cu}^\text{II}$ as Resource for Bipartite and Tripartite Quantum Entanglement and Coherence}

\author{Azadeh Ghannadan\orcidlink{0009-0005-1448-0919}}
\affiliation{Saeed's Quantum
	Information Group, P.O. Box 19395-0560, Tehran, Iran}
 
\author{Hamid Arian Zad\orcidlink{0000-0002-1348-1777}}
\email{Corresponding author: hamid.arian.zad@upjs.sk}
\affiliation{Department of Theoretical Physics and Astrophysics, Faculty of Science, P. J. S\v{a}f{\' a}rik University, Park Angelinum 9, 041 54 Ko\v{s}ice, Slovak Republic}

\author{Saeed Haddadi\orcidlink{0000-0002-1596-0763}}
\affiliation{Faculty of Physics, Semnan University, P.O. Box 35195-363, Semnan, Iran}
\affiliation{Saeed's Quantum
	Information Group, P.O. Box 19395-0560, Tehran, Iran}
 
\author{Jozef Stre{\v c}ka\orcidlink{0000-0003-1667-6841}}
\affiliation{Department of Theoretical Physics and Astrophysics, Faculty of Science, P. J. S\v{a}f{\' a}rik University, Park Angelinum 9, 041 54 Ko\v{s}ice, Slovak Republic}

\author{Zhirayr Adamyan}
\affiliation{Laboratory of Theoretical Physics, Yerevan State University, Alex Manoogian 1, 0025 Yerevan, Armenia}
\affiliation{CANDLE, Synchrotron Research Institute, 31 Acharyan Str., 0040 Yerevan, Armenia}

\author{Vadim Ohanyan\orcidlink{0000-0002-7810-7321}}
\affiliation{Laboratory of Theoretical Physics, Yerevan State University, Alex Manoogian 1, 0025 Yerevan, Armenia}
\affiliation{CANDLE, Synchrotron Research Institute, 31 Acharyan Str., 0040 Yerevan, Armenia}

\date{\today}

\begin{abstract}
We investigate key quantum characteristics of the mixed spin-(1/2,1,1/2) Heisenberg trimer under the influence of an external magnetic field. Specifically, we analyze the distributions of bipartite and tripartite entanglement quantified through the respective negativities, and the $l_1$-norm of coherence with the help of rigorous analytical and numerical methods. Our findings suggest that the heterotrinuclear molecular nanomagnet $[\{\text{Cu}^\text{II}\text{L}\}_2\text{Ni}^\text{II}(\text{H}_2\text{O})_2](\text{ClO}_4)_{2} . 3\text{H}_2\text{O}$, which represents an experimental realization of the mixed spin-(1/2,1,1/2) Heisenberg trimer, exhibits a significant bipartite entanglement between $\text{Cu}^\text{II}$ and $\text{Ni}^\text{II}$ magnetic ions along with robust tripartite entanglement among all three constituent $\text{Cu}^\text{II}\text{Ni}^\text{II}\text{Cu}^\text{II}$ magnetic ions. The significant bipartite and tripartite entanglement persists even at relatively high temperatures up to $37\,\text{K}$ and magnetic fields up to $46\,\text{T}$, whereby the coherence is maintained even at elevated temperatures. {It is evidenced that the aforementioned molecular complex with the magnetic core $\text{Cu}^\text{II}\text{Ni}^\text{II}\text{Cu}^\text{II}$ provides an intriguing quantum resource, which exhibits a star-shaped state within the singlet eigenstate at low magnetic fields and W-like state within the triplet eigenstate at moderate magnetic fields.} 
\end{abstract}

\maketitle


\section{Introduction}\label{sec:introduction}

Entanglement \cite{Horod2009}, a cornerstone of quantum mechanics, plays a crucial role in the study of molecular magnets \cite{Carlin1986,Kahn1993}, driving advancements in quantum information and condensed matter physics \cite{Arnesen2001,Amico2008,Benelli2015}. In this context, Heisenberg spin clusters serve as exemplary systems for exploring fundamental quantum properties, such as entanglement \cite{Tribedi2006,Troiani2011,str2020,Var2024} and quantum coherence \cite{Zurek2003,Kanai2022,Gassab2024}. These properties not only illuminate the intricate behavior of quantum systems but also pave the way for practical applications in quantum information \cite{Mahdavifar2022,Mahdavifar2024,Wang2001,Jaeger2018,Bao2020,Calixto2021}, magnetic resonance \cite{Sinha2003} and improving the sensor sensitivity \cite{Jensen2010,Sewell2012,Shen2020}. Our investigation is particularly inspired by the experimental developments in molecular magnets, which provide a tangible platform for testing theoretical predictions. By studying the quantum characteristics of Heisenberg spin clusters, we aim to deepen our understanding of these phenomena and their implications for future quantum technologies \cite{Amico2008, Horod2009}.
	
Different states coexist simultaneously through superposition in quantum systems, which is known as quantum coherence \cite{Baumgratz2014}. Many quantum technologies including cryptography and computing depend on this feature. It is well known that real systems can exhibit interference effects when they are coherent, which makes it easier to accurately control and manipulate their behavior. However, coherence is brittle and easily disrupted by interactions with the surroundings, leading to decoherence \cite{Schlosshauer2005,Schlosshauer2019}. As such, maintaining coherence is a significant obstacle to the development of useful quantum technologies. Generally speaking, the quantum coherence of real systems is affected by various factors, including environmental noise and general decoherence mechanisms \cite{Streltsov2017, mlhu2018}. Indeed, the environmental noise caused by some factors such as thermal fluctuations can disrupt the fragile quantum superpositions required for coherence.  Thus, studying quantum coherence is crucial for exploring the quantum characteristics of these systems under decoherence.
	
According to the hierarchy of quantum resources \cite{Ma2019,Ali2024}, quantum coherence and entanglement are different but related phenomena (quantum coherence $\supseteq$ entanglement), and they serve distinct purposes in quantum information processing.  Quantum coherence, as mentioned earlier, refers to the ability of a quantum system to exist in a superposition of states. This property is vital for some interesting tasks like quantum communication and quantum computation, where controlling and preserving coherence is crucial for performing complex calculations or transmitting information securely \cite{DiVincenzo1999,MLHu2012,Fan2015}. Entanglement, on the other hand, is a special kind of correlation that exists between the states of two or more quantum systems, where the states of each system are intrinsically linked, even when they are physically separated. In some cases, depending on the specific application, examining quantum coherence may be preferred over entanglement. Therefore, the reason to choose quantum coherence over entanglement (or vice versa) depends on the requirements of the considered quantum protocol or program, as well as specific constraints and resources.
	
{ In the present work, we investigate the quantum features of the mixed-spin (1/2,1,1/2) Heisenberg trimer. This model can} be experimentally realized by the heterotrinuclear molecular nanomagnet $[\{\text{Cu}^\text{II}\text{L}\}_2\text{Ni}^\text{II}(\text{H}_2\text{O})_2](\text{ClO}_4)_{2}  3\text{H}_2\text{O}$, hereafter referred to as the $\text{Cu}^\text{II}\text{Ni}^\text{II}\text{Cu}^\text{II}$ molecular complex. The magnetic properties of the $\text{Cu}^\text{II}\text{Ni}^\text{II}\text{Cu}^\text{II}$ complex characterized by a linear structure and a rather weak single-ion anisotropy were studied in Ref. \cite{Biswas2010}, where the strength of the exchange coupling between Cu$^{\rm II}$ and Ni$^{\rm II}$ ions was reported along with the single-ion anisotropy of the Ni$^{\rm II}$ ions.

In an earlier work \cite{Wang2005}, the magnetic properties of the $\text{Cu}^\text{II}\text{Ni}^\text{II}\text{Cu}^\text{II}$ complex were studied without considering single-ion anisotropy. Later,  Hari \emph{et al.} \cite{Hari2018} synthesized and characterized a $\text{Cu}^\text{II}\text{Ni}^\text{II}\text{Cu}^\text{II}$ complex with nonzero $\text{Cu}\cdots\text{Cu}$ exchange coupling, which was found to be much smaller than the $\text{Cu}\cdots\text{Ni}$ exchange couplings. They then investigated its magnetic properties. Comparing the results obtained for the magnetic behavior of the two structural models—one with and one without $\text{Cu}\cdots\text{Cu}$ exchange coupling—it was observed that the magnetization and magnetic susceptibility behavior did not change significantly. In a couple of recent theoretical works \cite{Vadim2024,Adamyan2024}, the ground-state phase diagram, bipartite entanglement, and magnetization properties of a trimeric model similar to that reported in Ref. \cite{Hari2018}, but with different Landé $g$-factors of Cu$^{\rm II}$ and Ni$^{\rm II}$ ions, have been investigated in detail.

Motivated by the results reported in Ref. \cite{Biswas2010}, we selected the $\text{Cu}^\text{II}\text{Ni}^\text{II}\text{Cu}^\text{II}$ complex without $\text{Cu}\cdots\text{Cu}$ exchange interaction to investigate the effects of the applied magnetic field and temperature on its quantum properties. This study provides a comprehensive examination of the mixed spin-(1/2,1,1/2) Heisenberg trimer model with nonzero single-ion anisotropy in the presence of an external magnetic field, offering novel quantum insights that have not been previously explored in this context.

The structure of the paper is as follows: In Sec. \ref{sec:model}, we present the model and derive its partition function from the exact solution. Section \ref{sec:DMandEnt} outlines the key formulas and theoretical approaches used in our analysis of the global density matrix of the system, as well as two reduced density matrices for a nonzero value of the single-ion anisotropy of the spin integer particle. This section also explores the tripartite entanglement negativity and two bipartite negativities with respect to the temperature and magnetic field. In Sec. \ref{sec:coherence}, we study the quantum coherence of the system and compare it with the entanglement negativities. In Sec. \ref{sec:Exp}, we present a theoretical prediction for the degree of the bipartite and tripartite entanglement, as well as coherence of the heterotrinuclear molecular nanomagnet $\text{Cu}^\text{II}\text{Ni}^\text{II}\text{Cu}^\text{II}$, which represents an experimental realization of the mixed spin-(1/2,1,1/2) Heisenberg trimer. Finally, Sec. \ref{sec:conclusions} provides a summary and outlook.

\section{ Model}\label{sec:model}
Let us consider the mixed spin-(1/2,1,1/2) Heisenberg trimeric cluster, which is experimentally motivated by the heterotrinuclear $\text{Cu}^\text{II}\text{Ni}^\text{II}\text{Cu}^\text{II}$ complex.
{ The model can be described by the following Hamiltonian:}
\begin{eqnarray}\label{Eq:hamiltonian}
\!\hat{H}\!\!=\!\! J\big(\hat{{\boldsymbol s}}_\text{a} \!\cdot\! \hat{{\boldsymbol S}}_\text{b} + \hat{{\boldsymbol S}}_\text{b} \!\cdot\! \hat{{\boldsymbol s}}_\text{c} \big) \!\!+\!\! D\big(\hat{{S}}_\text{b}^{z}\big)^{2} \!\!\!-\! g \mu_{\text{B}} B \big(\hat{{s}}_\text{a}^{z} + \hat{{S}}_\text{b}^{z} + \hat{{s}}_\text{c}^{z} \big).
\nonumber \\
\end{eqnarray}
Here, the first term of the Hamiltonian (\ref{Eq:hamiltonian}) expresses the nearest-neighbor exchange interactions between the spin operators $\hat{{\boldsymbol s}}_{\text{a}}$ and $\hat{{\boldsymbol s}}_{\text{c}}$ { assigned to $\text{Cu}^\text{II}$ ions and the spin operator $\hat{{\boldsymbol S}}_{\text{b}}$ assigned to the $\text{Ni}^\text{II}$ ion respectively,} with the antiferromagnetic coupling constant $J > 0$. { The second term, $D$, refers to the single-ion anisotropy of the $\text{Ni}^\text{II}$ ion. Finally, the last term is a standard Zeeman term, where $g$ and $B$ represent the Landé $g$-factor and the static magnetic field for the $z$-axis and $\mu_\text{B}$ is the Bohr magneton.} The Hamiltonian  (\ref{Eq:hamiltonian}) can be solved analytically through the exact diagonalization method and a full set of eigenvalues and eigenvectors can be obtained. { The twelve energy eigenvalues, corresponding eigenvectors and the respective probability amplitudes of the eigenvectors, written in the standard basis of the eigenvectors of $z$-component of all three spins $\rvert \phi \rangle = \rvert s_{\text{a}}^{z},S_{\text{b}}^{z},s_{\text{c}}^{z}\rangle$, are given in Table \ref{tab:table1} and Table \ref{tab:table2} reported in Appendix \ref{Sec:appendix a}.} Next, the Gibbs free energy of the system can be calculated using the following expression:
\begin{eqnarray}\label{Eq:free_energy}
\mathcal{G} = - k_{\text{B}} T ~\ln Z,
\end{eqnarray}
where $Z$ is the partition function of the system that can be explicitly obtained from the sum of the exponentials of the eigenenergies as:
\begin{equation}\label{Eq:part_func}
\begin{split}
&Z = \sum_{i = 1}^{12} \exp (-\beta E_{i}) = 1+ 2 ~\cosh (\beta h) \exp (-\beta D)  \\
&+\exp \left(-\!\beta (D - J)\right) \!+\! 2~ \cosh \left( 2\beta h \right) \exp \left(-\beta(D+J)\right) \\
&+4~ \cosh \left(\beta h\right) \exp \left(-\frac{\beta D}{2}\right) \cosh \left(\frac{\beta}{2}\sqrt{D^2+4J^2}\right)  \\
& + 2 ~\cosh \left(\frac{\beta}{2}\sqrt{(D-J)^2+8J^2}\right) \exp \left(-\frac{\beta}{2}(D-J)\right).
\end{split}
\end{equation}
Here, $\beta = 1/(k_{\text{B}}T)$, where $k_{\text{B}}$ is the Boltzmann constant and $T$ is the absolute temperature. For convenience, we introduced a new parameter \( h = g \mu_{\text{B}} B \) representing the Zeeman term related to the applied magnetic field. Using the Gibbs free energy (\ref{Eq:free_energy}), we can derive other quantities such as magnetization, { given by:}
\begin{equation}\label{Eq}
M = - \frac{\partial \mathcal{G}}{\partial B}.
\end{equation}

\section{The density matrices and entanglement measures}\label{sec:DMandEnt}
Having accomplished the eigenvalue problem of the mixed spin-(1/2,1,1/2) Heisenberg trimer and having the partition function in hand, one can step forward to evaluation of the density matrix which allows the calculation of bipartite and tripartite entanglement via a measure of negativity \cite{Vid-Wer}. As the first step, the overall density operator is defined by the formula:
\begin{eqnarray}\label{Eq:tot_den_mat}
\hat{\rho} = \frac{1}{Z} \exp (-\beta \hat{H})= \frac{1}{Z} \sum_{i = 1}^{12} \exp (- \beta E_{i})\rvert \psi_{i} \rangle \langle \psi_{i} \lvert,
\end{eqnarray}
{ and the density matrix corresponding to the density operator is written in the standard basis $\rvert \phi \rangle$ of eigenvectors of $z$-components of all three constituent spins, reported in Appendix \ref{Sec:appendix b}.}

Then, to evaluate the bipartite entanglement between any spin pair, one should follow the calculation of the reduced density operator related to the spin pair by tracing out the degrees of freedom of the remaining spin of the mixed spin-(1/2,1,1/2) Heisenberg cluster. { Although the reduced density operators  $\hat{\varrho}_{\text{ab}}$ and  $\hat{\varrho}_{\text{bc}}$ are not equivalent, the results of the negativity (as well as coherence) related to these subsystems are identical. Therefore, hereafter the following bipartite state will be only considered for calculating the entanglement and coherence between parties $s_{\text{a}}$ and $S_{\text{b}}$, given by:}
\begin{equation}\label{Eq:den_mat12}
\hat{\varrho}_{\text{ab}} \!\!=\! \text{Tr}_{s_{\text{c}}} \hat{\rho} \!=\!\! \frac{1}{Z} \! \sum_{i = 1}^{12}\! \sum_{s_{\text{c}}^{z}= \pm 1/2} \!\!\!\!\!\text{exp}(- \!\beta E_{i})\langle s_{\text{c}}^{z}\rvert \psi_{i} \rangle \langle \psi_{i} \lvert s_{\text{c}}^{z} \rangle .
\end{equation}
In a similar way, the reduced density operator relevant to { two spin-$1/2$ entities} $s_{\text{a}}$ and $s_{\text{c}}$ is calculated by tracing out the degrees of freedom of { the spin-$1$ entity} $S_{\text{b}}$:
\begin{equation}\label{Eq:den_mat13}
\hat{\varrho}_{\text{ac}} \!\!=\!\! \text{Tr}_{S_{\text{b}}} \hat{\rho} \!=\!\! \frac{1}{Z} \! \sum_{i = 1}^{12} \!\sum_{S_{\text{b}}^{z}=0, \pm 1} \!\!\!\!\!\text{exp}(-\! \beta E_{i})\langle\! S_{\text{b}}^{z}\rvert \psi_{i} \rangle \langle \psi_{i} \lvert S_{\text{b}}^{z} \rangle .
\end{equation}

The next step is to write the matrix form of the reduced density operators and partially transpose them with respect to one of the spins of their basis. The elements of the partially transposed density matrices $\varrho_{\text{ab}}$ and $\varrho_{\text{ac}}$ can be written as follows:
\begin{equation}\label{Eq:red_den_mat}
\begin{split}
&\langle s_{\text{a}}^{z_{'}},S_{\text{b}}^{z_{'}}\lvert \varrho_{\text{ab}} \rvert s_{\text{a}}^{z},S_{\text{b}}^{z} \rangle^{\text{T}_{\text{a}}} \!=\! \langle s_{\text{a}}^{z},S_{\text{b}}^{z_{'}}\lvert \varrho_{\text{ab}} \rvert s_{\text{a}}^{z_{'}},S_{\text{b}}^{z}\rangle, \\
&\langle s_{\text{a}}^{z_{'}},s_{\text{c}}^{z_{'}}\lvert \varrho_{\text{ac}} \rvert s_{\text{a}}^{z},s_{\text{c}}^{z} \rangle^{\text{T}_{\text{c}}} \!=\! \langle s_{\text{a}}^{z_{'}},s_{\text{c}}^{z}\lvert \varrho_{\text{ac}} \rvert s_{\text{a}}^{z},s_{\text{c}}^{z_{'}}\rangle.
\end{split}
\end{equation}

Ultimately, the bipartite negativities ${N}_{\text{ab}}$ and ${N}_{\text{ac}}$ measuring the magnitude of entanglement between the spin pairs $s_{\text{a}}$-$S_{\text{b}}$ and $s_{\text{a}}$-$s_{\text{c}}$ are calculated according to the formula by Vidal and Werner \cite{Vid-Wer} that originates from the Peres-Horodecki separability criterion \cite{Peres1996}:
\begin{eqnarray}\label{Eq:Neg12}
{N}_{\text{ab}} = \sum_{i = 1}^{6} \frac{|\lambda_{i}| - \lambda_{i}}{2}, \hspace{0.5cm}  {N}_{\text{ac}} = \sum_{i = 1}^{4} \frac{|\lambda_{i}| - \lambda_{i}}{2},
\end{eqnarray}
where $\lambda_{i}$ are the eigenvalues of the respective partially transposed density matrices.
{ Both of the negativities $ N_{\text{ab}}$ and ${N}_{\text{ac}}$ are calculated analytically and all eigenvalues that contribute to $ N_{\text{ab}}$ and ${N}_{\text{ac}}$ are presented in Appendix \ref{Sec:appendix c}.}

The tripartite entanglement of the mixed spin-(1/2,1,1/2) Heisenberg cluster is quantified through the tripartite negativity, which is calculated from the three different partial transposes of the overall density matrix. The first partial transpose of the overall density matrix with respect to the first spin of the basis $\rho^{\text{T}_{\text{a}}}$ is performed by fixing the quantum spin numbers $S_{\text{b}}^{z}$ and $s_{\text{c}}^{z}$ in the basis $\rvert s_{\text{a}}^{z}, S_{\text{b}}^{z}, s_{\text{c}}^{z}\rangle$ and interchanging the quantum spin number $s_{\text{a}}^{z}$.
The same procedure is repeated to obtain the partial transpose of the overall density matrix with respect to the second and third spins of the basis:
\begin{equation} \label{redden}
\begin{split}
\langle s_{\text{a}}^{z_{'}},S_{\text{b}}^{z_{'}},s_{\text{c}}^{z_{'}}\lvert \rho \rvert s_{\text{a}}^{z},S_{\text{b}}^{z},s_{\text{c}}^{z} \rangle^{\text{T}_{\text{a}}} \!=\! \langle s_{\text{a}}^{z},S_{\text{b}}^{z_{'}},s_{\text{c}}^{z_{'}}\lvert \rho \rvert s_{\text{a}}^{z_{'}},S_{\text{b}}^{z},s_{\text{c}}^{z}\rangle. \\
\langle s_{\text{a}}^{z_{'}},S_{\text{b}}^{z_{'}},s_{\text{c}}^{z_{'}}\lvert \rho \rvert
 s_{\text{a}}^{z},S_{\text{b}}^{z},s_{\text{c}}^{z} \rangle^{\text{T}_{\text{b}}} \!=\! \langle s_{\text{a}}^{z_{'}},S_{\text{b}}^{z},s_{\text{c}}^{z_{'}}\lvert \rho \rvert s_{\text{a}}^{z},S_{\text{b}}^{z_{'}},s_{\text{c}}^{z}\rangle, \\
\langle s_{\text{a}}^{z_{'}},S_{\text{b}}^{z_{'}},s_{\text{c}}^{z_{'}}\lvert \rho \rvert s_{\text{a}}^{z},S_{\text{b}}^{z},s_{\text{c}}^{z} \rangle^{\text{T}_{\text{c}}} \!=\! \langle s_{\text{a}}^{z_{'}},S_{\text{b}}^{z_{'}},s_{\text{c}}^{z}\lvert \rho \rvert s_{\text{a}}^{z},S_{\text{b}}^{z},s_{\text{c}}^{z_{'}}\rangle.
\end{split}
\end{equation}
The bipartite negativities that are calculated from the partially transposed density matrices \eqref{redden} are respectively denoted by ${N}_{\text{a-bc}}$, ${N}_{\text{b-ac}}$ and ${N}_{\text{c-ab}}$ such that ${N}_{\text{I-JK}}=\sum_{i=1}^{12} \frac{|\lambda_{i} | - \lambda_{i}}{2}$ with $\text{I}=\text{a,b,c}$ and $\text{JK}=\text{bc}, \text{ac}, \text{ab}$. 
Consequently, the tripartite negativity of the mixed spin-(1/2,1,1/2) Heisenberg trimer can be calculated as the geometric mean of the bipartite negativities \cite{Sabin2008}:
\begin{equation}\label{Eq:Neg}
{N}_{\text{abc}} = \sqrt[3]{{N}_{\text{a-bc}}~ {N}_{\text{b-ac}}~ {N}_{\text{c-ab}}}~.
\end{equation}
Note that the matrix forms of overall density matrix, reduced density matrices, and their partially transposed density matrices are presented in Appendix \ref{Sec:appendix a}.

\begin{figure}[t]
\centering
\resizebox{0.48\textwidth}{!}{
\includegraphics[trim = 0 0 0 0, clip]{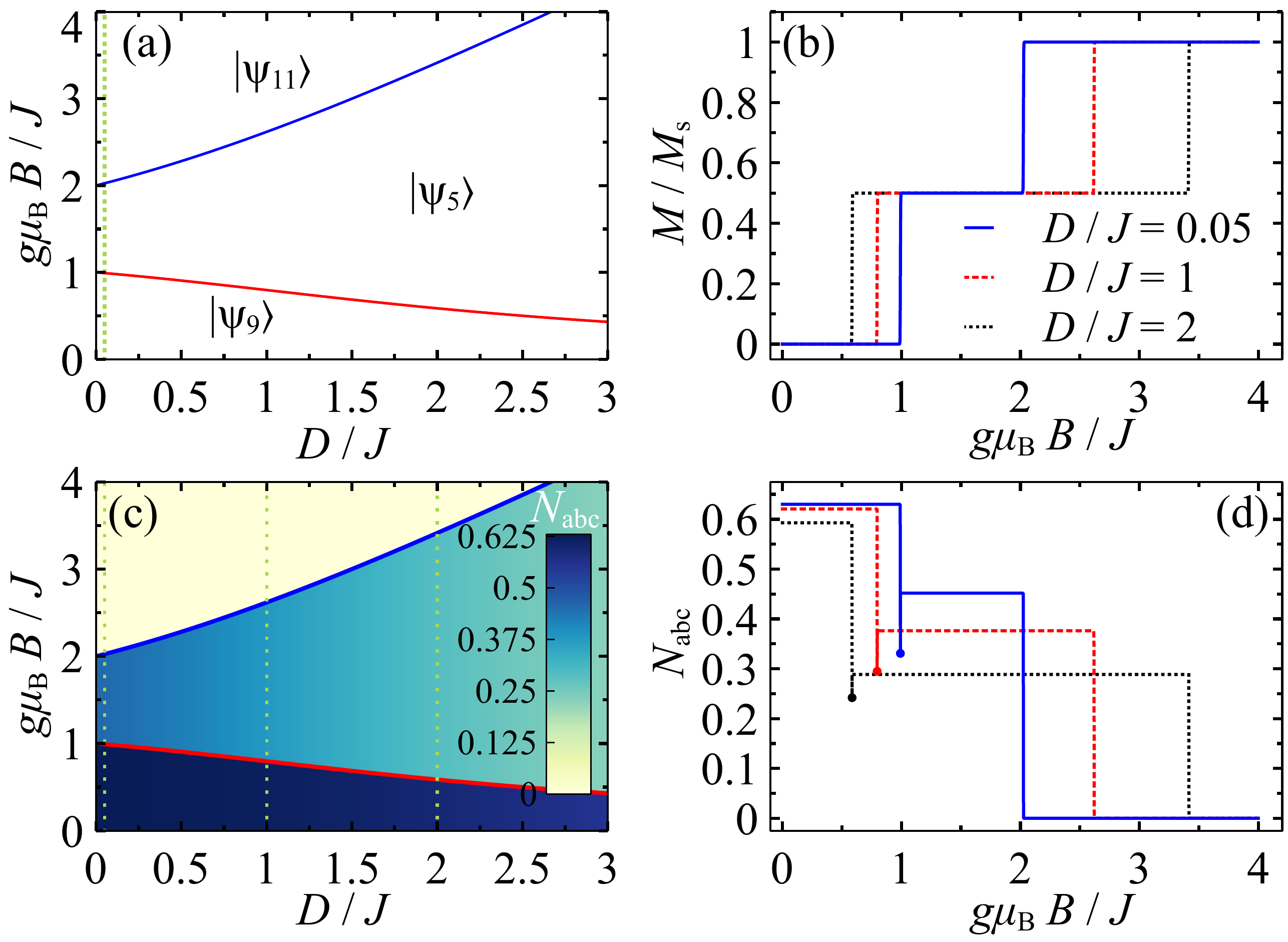}}
\caption{(a) The ground-state phase diagram of the mixed spin-(1/2,1,1/2) Heisenberg trimer in the parameter space $D / J$ vs. $g \mu_{\text{B}} B / J$; (b) The zero-temperature magnetization normalized with respect to the saturation value ($M_\text{s} = 2g\mu_\text{B}$) as a function of the magnetic field for three different values of the single-ion anisotropy $D/J$; (c) The zero-temperature density plot of the tripartite negativity ${N}_\text{abc}$ in the $g\mu_\text{B}B/J - D/J$ plane. { Vertical dotted lines indicate fixed values of $D/J = \{0.05, 1, 2\}$}; (d) The tripartite negativity $N_{\text{abc}}$ as a function of the magnetic field for three different values of $D/J$ at zero temperature.}
\label{fig:GSPD_TriNeg_Mag}
\end{figure}

Now, let us begin with a detailed analysis of the ground-state behavior of the mixed spin-(1/2,1,1/2) Heisenberg trimer in presence of a magnetic field. To proceed to the results, all interaction terms involved in the Hamiltonian \eqref{Eq:hamiltonian} will be normalized with respect to the antiferromagnetic coupling constant $J>0$, which will henceforth serve as an energy unit. The ground-state phase diagram of the mixed spin-(1/2,1,1/2) Heisenberg trimer is presented in Fig. \ref{fig:GSPD_TriNeg_Mag}(a) in the parameter space $D / J$ vs. $g \mu_{\text{B}} B / J$. It turns out that there are three different ground states unambiguously given by the eigenvectors $\rvert\psi_9\rangle$, $\rvert\psi_5 \rangle$ and $\rvert\psi_{11}\rangle$ specifically quoted in Appendix \ref{Sec:appendix a}. The inspection of energy eigenvalues reveals the following phase boundaries between three available ground states $\rvert\psi_9\rangle$, $\rvert\psi_5 \rangle$ and $\rvert\psi_{11}\rangle$:
\begin{equation}
	\begin{split}
		&\rvert \psi_5 \rangle - \rvert \psi_9 \rangle : h \!=\! \frac{J}{2} \!+\! \frac{1}{2} \left(\sqrt{(D \!-\! J)^2 \!+\! 8 J^2 } \!-\! \sqrt{D^2 \!+\! 4 J^2} \right) \\
		&\rvert \psi_{11} \rangle - \rvert \psi_5 \rangle :   h = J+\frac{D}{2} +\frac{1}{2} \sqrt{D^2+4J^2}
	\end{split}
\end{equation}

In Fig. \ref{fig:GSPD_TriNeg_Mag}(b), the zero-temperature magnetization curves displaying zero, one-half and saturation plateaus are illustrated at three different cuts of the relevant ground-state phase diagram. At low magnetic fields, the singlet ground state $\rvert \psi_9 \rangle$ having character of a quantum superposition of four basis states $\rvert \downarrow 1 \downarrow \rangle$, $\rvert \uparrow -1 \uparrow \rangle$, $\rvert \uparrow 0 \downarrow \rangle$, and $\rvert \downarrow 0 \uparrow \rangle$,  with zero total spin moment $S_\text{tot}^{z} = 0$ is favored. This ground state corresponds to a zero magnetization plateau. At moderate magnetic fields, one contrarily encounters the triplet ground state $\rvert \psi_5 \rangle$ with character of a quantum superposition of three basis states
$\rvert \uparrow 1 \downarrow \rangle$, $\rvert \downarrow 1 \uparrow \rangle$, $\rvert \uparrow 0 \uparrow \rangle$ with the total spin moment $S_\text{tot}^{z} = 1$. The ground state $\rvert \psi_5 \rangle$ thus corresponds to an intermediate one-half magnetization plateau being stable in a rather wide region of the magnetic fields. Finally, one detects the fully saturated ground state $\rvert \psi_{11}\rangle = \rvert \uparrow 1 \uparrow \rangle$, which is reached at high enough magnetic fields where all three spins are fully aligned with the magnetic-field direction.

Figure \ref{fig:GSPD_TriNeg_Mag}(c) displays the density plot of the tripartite negativity $N_{\text{abc}}$ for the mixed spin-(1/2,1,1/2) Heisenberg trimer in the same parameter region as used in Fig. \ref{fig:GSPD_TriNeg_Mag}(a). As expected, the tripartite negativity is highest in the ground state $\rvert \psi_9 \rangle$ reaching a maximum value of $N_{\text{abc}} = 0.625$ at zero single-ion anisotropy. The singlet ground state $\rvert \psi_9 \rangle$ thus exhibits the most pronounced collective quantum features. In contrast, the collective quantum features are somewhat reduced within another quantum ground state $\rvert \psi_5 \rangle$ due to partial spin alignment towards the magnetic field realized in this one-half plateau state. Consequently, the tripartite negativity is substantially lower in the triplet ground state $\rvert \psi_5 \rangle$ in comparison with the singlet ground state $\rvert \psi_9 \rangle$. Finally, the tripartite negativity becomes zero in the fully polarized ground state $\rvert \psi_{11} \rangle$. 

To gain better overall insight, the magnetic field variations of the tripartite negativity are shown in Fig. \ref{fig:GSPD_TriNeg_Mag}(d) for three different cuts of the relevant density plot. A prominent feature of the easy-plane single-ion anisotropy $D>0$ consist in the reduction of the tripartite negativity. This effect is more pronounced for the triplet ground state \( |\psi_5\rangle \) compared to the singlet ground state \( |\psi_9\rangle \). Notably, one also detects a substantial drop in the tripartite negativity directly at a coexistence point of the ground states \( |\psi_9\rangle \) and \( |\psi_5\rangle \) visualized in Fig. \ref{fig:GSPD_TriNeg_Mag}(d) by a solid circle. This striking phenomenon arises from a mixed state realized exactly at the ground-state boundary, which can be characterized through the density operator \( \rho_\text{mix} = \frac{1}{2}(|\psi_5\rangle \langle \psi_5| + |\psi_9\rangle \langle \psi_9|) \). The mixedness of this ground state is responsible for a significant reduction in the tripartite entanglement, which is eventually lowered below the value corresponding to the triplet ground state \( |\psi_5\rangle \) with a weaker tripartite entanglement. The similar finding was reported in Ref. \cite{Vargova2023} for the mixed spin-(1/2,1) Heisenberg tetramer.

\begin{center}
	\begin{figure}[H]
		\centering
		\includegraphics[scale=0.29]{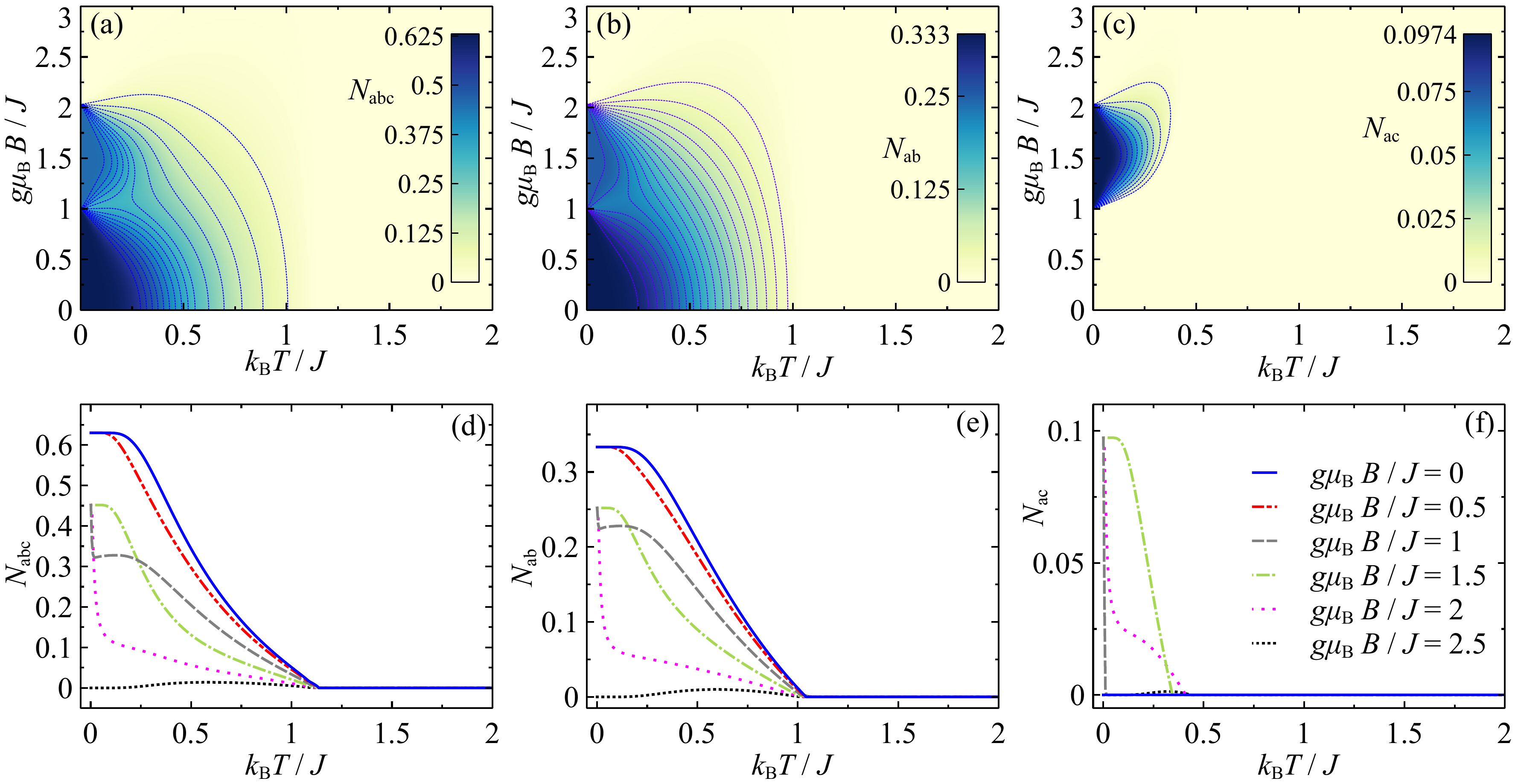} 
		\caption{(a) The tripartite negativity $N_{\text{abc}}$, (b) the bipartite negativity $N_\text{ab}$, and (c) the bipartite negativity $N_\text{ac}$ of the mixed spin-(1/2,1,1/2) Heisenberg trimer in the $g\mu_\text{B}B/J - k_\text{B}T/J$ plane by assuming $D/J=0.05$; (d) Temperature dependencies of the $N_{\text{abc}}$, (e) $N_\text{ab}$ and (f) $N_\text{ac}$ for a few selected values of the magnetic field and $D/J=0.05$.}
		\label{fig:TriBipNeg}
	\end{figure}
\end{center}

To provide a clear understanding of the thermal behavior of the mixed spin-(1/2,1,1/2) Heisenberg trimer, the density plot of tripartite as well as bipartite negativities are depicted in Fig. \ref{fig:TriBipNeg}(a)-(c) for the particular case with the fixed value of the uniaxial single-ion anisotropy $D /J = 0.05$. 
{ The selection of a relatively weak easy-plane single-ion anisotropy for the results presented in Fig. \ref{fig:TriBipNeg} and all subsequent figures is inspired by previous experimental findings for the compound \( \text{Cu}^{\text{II}}\text{Ni}^{\text{II}}\text{Cu}^{\text{II}} \) \cite{Biswas2010}. This parameter will be accordingly not treated as a free modeling variable. A more comprehensive study, which will include consideration of a wider range of easy-axis and easy-plane single-ion anisotropies as well as the effects of magnetic fields applied in different directions, is planned for future work.} 

\onecolumngrid
\twocolumngrid

It is quite clear that the magnetic-field-driven variations of the tripartite negativity $N_{\text{abc}}$ at sufficiently low temperatures are in a perfect agreement with the abrupt changes of the tripartite negativity detected at zero temperature around the magnetic fields $g\mu_{\text{B}} B/J \approx 1$ and $2$ [cf. Fig. \ref{fig:TriBipNeg}(a),(d)]. 
We note that, the tripartite negativity ranges from 0 to 1, where for our case its maximum is 0.625. It can be inferred from Figs. \ref{fig:TriBipNeg}(b) and \ref{fig:TriBipNeg}(c) that the bipartite negativities ${N}_{\text{ab}}$ and ${N}_{\text{ac}}$ form at low enough temperatures plateaus quite similarly as the tripartite negativity $N_{\text{abc}}$ does. The tripartite negativity \( N_{\text{abc}} \) and the bipartite negativity \( N_{\text{ab}} \) display their highest achievable values, \( N_{\text{abc}} = 0.625 \) and \( N_{\text{ab}} = 1/3 \), in a low-field region inherent to the singlet ground state $\rvert \psi_9 \rangle$, whereas the maximum of bipartite negativity ${N}_{\text{ac}}\approx 0.0974$ can be found at moderate magnetic fields supporting the triplet ground state $\rvert \psi_5 \rangle$. Generally, the maximum values of both bipartite negativities are restricted to between 0 and 0.5. Absence of a direct exchange interaction between the outer spins $s_{\text{a}}$ and $s_{\text{c}}$ generally results in a lower value of the bipartite entanglement ${N}_{\text{ac}}$ compared to the ${N}_{\text{ab}}$ one assigned to directly interacting central and outer spins. A gradual vanishing of tripartite and bipartite negativities is observed along the temperature axis in all three density plots shown in upper panel.

The bottom panel in Fig. \ref{fig:TriBipNeg} demonstrates typical temperature dependencies of the tripartite and bipartite negativities $N_{\text{abc}}$, ${N}_{\text{ab}}$ and ${N}_{\text{ac}}$ for a few selected values of the magnetic field. A common property of all temperature dependencies of tripartite and bipartite negativities in Fig. \ref{fig:TriBipNeg}(d)-(f) is the gradual thermally-induced decline in the tripartite and bipartite entanglement. In Fig. \ref{fig:TriBipNeg}(d) the tripartite negativity starts from its highest value $N_{\text{abc}} = 0.625$ achieved at zero magnetic field and vanishes at $k_{\text{B}} T / J \approx 1.15$. As the magnetic field strengthens, the initial value of the tripartite negativity gradually decreases, while the threshold temperature at which it becomes zero remains unchanged. Contrary to this, the negativity ${N}_{\text{ac}}$ measuring the bipartite entanglement between the outer spins is zero at low magnetic fields, but starts from the initial value ${N}_{\text{ac}} \approx 0.0974$ at moderate magnetic fields $g \mu_{\text{B}} B/J \gtrsim 1$ and is rapidly suppressed upon increasing temperature terminating at the threshold temperature $k_{\text{B}} T / J \approx 0.4$ [see Fig. \ref{fig:TriBipNeg}(f)]. The bipartite negativity ${N}_{\text{ab}}$ follows trends similar to those of the tripartite negativity $N_{\text{abc}}$ when reaching a maximum initial value ${N}_{\text{ab}} = 1/3$ and vanishing at the same threshold temperature $k_{\text{B}} T / J \approx 1.1$ [see Fig. \ref{fig:TriBipNeg}(e)]. A less pronounced reentrance in the tripartite and bipartite entanglement
can be found at magnetic fields slightly exceeding the saturation value $g\mu_{\text{B}} B/J \gtrsim 2.5$ due to thermal excitations towards higher-energy levels. Besides, the peculiar changes in tripartite and bipartite entanglement observable around the magnetic fields $g\mu_{\text{B}}B/J \approx 1$ and $2$ can be again attributed to the relevant magnetic-field-driven phase transitions.

\onecolumngrid
\begin{center}
	\begin{figure}[H]
		\centering
			\includegraphics[scale=0.29]{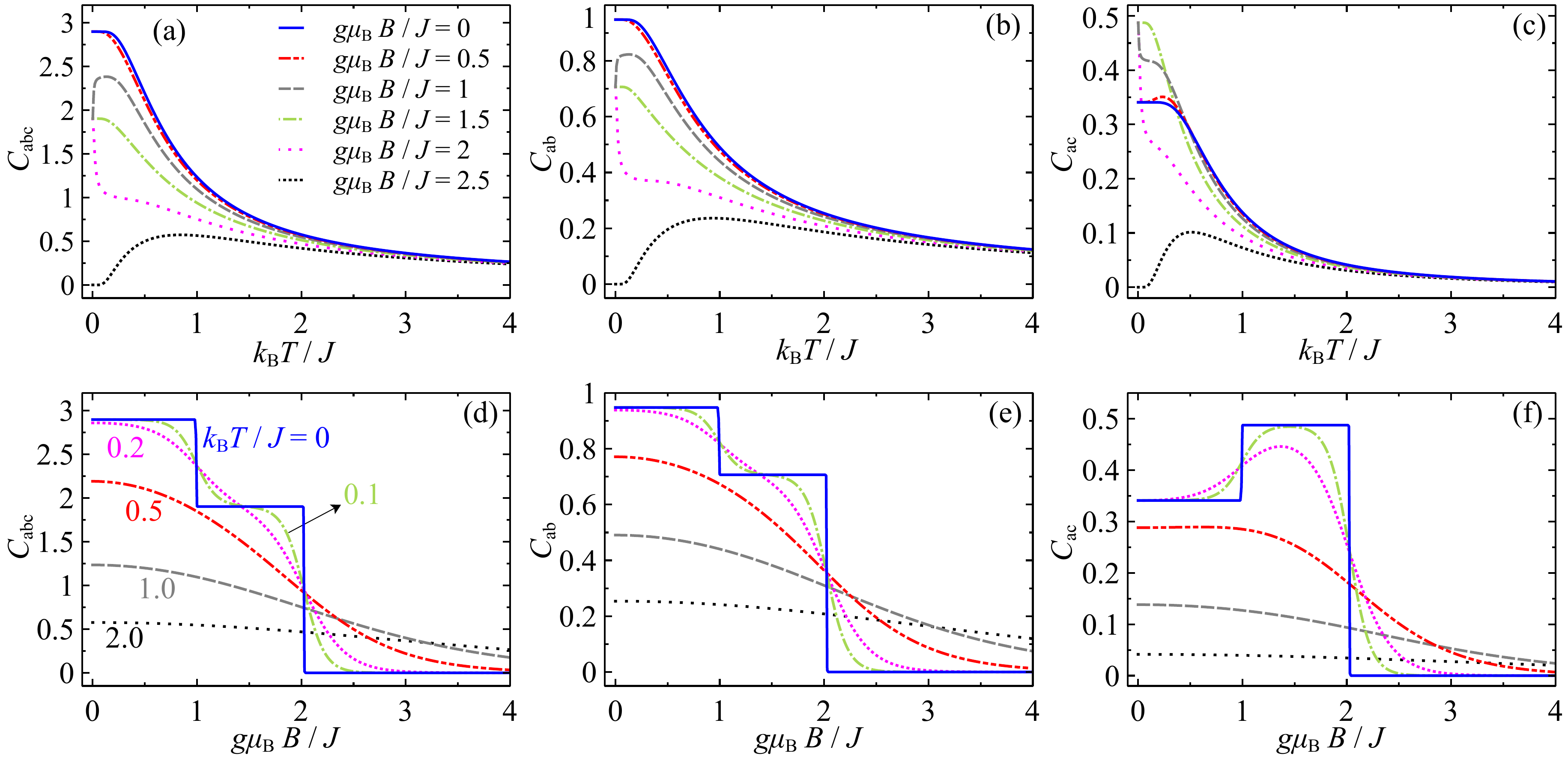}
		\caption{(a)-(c) The temperature dependencies of the quantum coherence  (tripartite and bipartite) of the mixed spin-(1/2,1,1/2) Heisenberg trimer for a few selected values of the magnetic field and one fixed value of the single-ion anisotropy $D/J=0.05$, (d)-(f)  the quantum coherence as a function of the magnetic field for the fixed value of the single-ion anisotropy $D/J=0.05$ and { six selected temperatures.}
		} \label{fig:QC}
	\end{figure}
\end{center}
\twocolumngrid

\section{Quantum coherence}\label{sec:coherence}
There are several ways to evaluate quantum coherence \cite{Streltsov2017}. Among these, the $l_{1}$-norm of quantum coherence, which is widely used in quantum physics, has the following definition:
\begin{equation}\label{Quantumcoherence}
 C=\sum_{i\neq j} |\rho_{i,j}|.
\end{equation}
According to Eq. \eqref{Quantumcoherence}, the $l_{1}$-norm of coherence can be expressed as the sum of absolute values of the off-diagonal elements corresponding to the selected basis \cite{mlhu2018}.

To find more details regarding the quantum resources in the system under consideration, we have plotted the $l_1$-norm of quantum coherence for the tripartite ($C_{\text{abc}}$) and bipartite ($C_{\text{ab}}$ and $C_{\text{ac}}$) states in Fig. \ref{fig:QC}. As can be seen in Fig. \ref{fig:TriBipNeg}, the tripartite and bipartite entanglement disappears at a certain threshold temperature, however, the quantum coherence of tripartite and bipartite states captured by $l_1$-norm of coherence \eqref{Quantumcoherence} remains nonzero even at higher temperatures as illustrated in Fig. \ref{fig:QC}(a)-(c)
for a few selected values of the magnetic field and $D/J=0.05$. This result is expected because increasing temperature can reduce quantum correlations, such as entanglement, due to rising role of thermal fluctuations in the system. The quantum coherence can nevertheless remain a significant resource compared to entanglement based on the hierarchy of quantum resources \cite{Ma2019} as the amount of entanglement is bounded by the quantum coherence \cite{Streltsov2015}. It is worth noting that the qualitative behavior of quantum coherence is similar to that of entanglement, apart from some minor differences [compare Figs. \ref{fig:TriBipNeg}(d)-(f) and \ref{fig:QC}(a)-(c)]. Moreover, despite the quantitative differences between tripartite and bipartite quantum coherences, their qualitative behavior is often the same, except for the bipartite coherence $C_{\text{ac}}$ at low temperatures.

Figure \ref{fig:QC}(d)-(f) shows the thermal coherence of the mixed spin-(1/2,1,1/2) Heisenberg trimer as a function of the magnetic field $g \mu_{\text{B}} B/J$ for six selected values of temperature and one fixed value of the single-ion anisotropy $D/J=0.05$. Regardless of temperature and single-ion anisotropy effects, one can see that the tripartite and bipartite quantum coherences decrease upon strengthening of the magnetic field. { The only exception to this rule is the $l_1$-norm of the bipartite coherence $C_{\text{ac}}$ [see Fig. \ref{fig:QC}(f)], which displays a peculiar transient strengthening due to the magnetic field at low temperatures. By inspecting Fig. \ref{fig:TriBipNeg}(c), we observe that the bipartite quantum entanglement $N_{\text{ac}}$ emerges within the field interval $1<g \mu_{\text{B}} B/J<2$ associated to the ground state $|\psi_5\rangle$ that results in an increase of bipartite coherence in this range.} In general, the suppression of thermal coherence by the magnetic field can be related to the Zeeman effect, which causes splitting of energy levels into sublevels (i.e. splitting of spin multiplets) and energetically favors the sublevels with the highest value of $z$-component of the total spin. As a result, the qualitative behavior of tripartite coherence [Fig. \ref{fig:QC}(d)] at zero and low temperatures is quite similar to the tripartite entanglement and magnetization [see blue lines in Figs. \ref{fig:GSPD_TriNeg_Mag}(b) and \ref{fig:GSPD_TriNeg_Mag}(d)]. The inclusion of a relatively small single-ion anisotropy may further cause zero-field splitting of energy levels and hence, it may additionally enhance the effect of external magnetic field. Altogether, the combined effect of magnetic field and single-ion anisotropy results in a more rapid loss of quantum coherence of the mixed spin-(1/2,1,1/2) Heisenberg trimer under thermal conditions. While the tripartite coherence and tripartite entanglement disappear at absolute zero temperature at the saturation field, it can be observed from Figs. \ref{fig:TriBipNeg} and \ref{fig:QC} that increasing temperature can delay this disappearance for larger magnetic fields.

\section{Quantum features of molecular nanomagnet $\text{Cu}^\text{II}\text{Ni}^\text{II}\text{Cu}^\text{II}$}
\label{sec:Exp}

 It has been verified in Ref. \cite{Biswas2010} that magnetic properties of the molecular compound $\text{Cu}^\text{II}\text{Ni}^\text{II}\text{Cu}^\text{II}$ can be accurately modeled by the mixed spin-(1/2,1,1/2) Heisenberg trimer with an isotropic exchange-coupling constant \( J = 22.8\,\text{cm}^{-1} \), a small nonzero single-ion anisotropy \( D = 0.05\,\text{cm}^{-1} \), and an average gyromagnetic ratio \( g = 2.227 \) for $\text{Cu}^\text{II}$ and $\text{Ni}^\text{II}$ magnetic ions \cite{Biswas2010}. In the following, we will adopt this set of parameters to make theoretical predictions for the tripartite negativity and coherence of the  heterotrinuclear molecular complex $\text{Cu}^\text{II}\text{Ni}^\text{II}\text{Cu}^\text{II}$.

\begin{figure}[hbt!]
	\centering
		\includegraphics[scale=0.33]{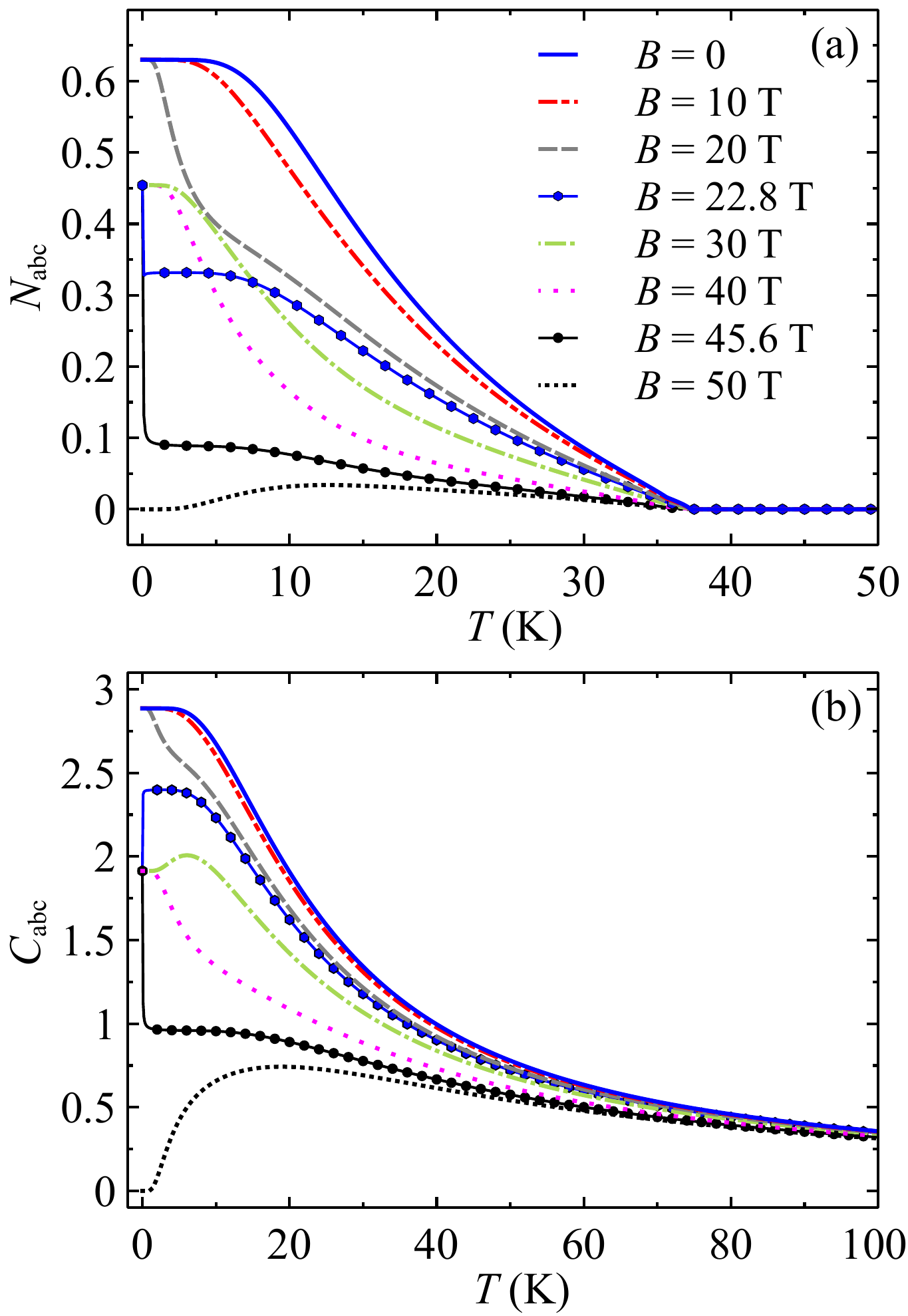} 
	\caption{Temperature variations of the tripartite negativity \( N_\text{abc} \) (a) and the quantum coherence \( C_\text{abc} \) (b) of the mixed spin-(1/2,1,1/2) Heisenberg trimer with an exchange-coupling constant \( J = 22.8\,\text{cm}^{-1} \), a single-ion anisotropy \( D = 0.05\,\text{cm}^{-1} \), and gyromagnetic ratio \( g = 2.227 \) adjusted according to Ref. \cite{Biswas2010} to a theoretical modeling of the molecular nanomagnet $\text{Cu}^\text{II}\text{Ni}^\text{II}\text{Cu}^\text{II}$. The labels given in the panel (a) determine the considered magnetic-field strengths.}
\label{fig:exp_Trip_Neg_Coh}
\end{figure}

Figure \ref{fig:exp_Trip_Neg_Coh}(a) illustrates temperature dependencies of the tripartite negativity \( N_\text{abc}\) of the mixed spin-(1/2,1,1/2) Heisenberg trimer adjusted to a theoretical modeling of the molecular nanomagnet $\text{Cu}^\text{II}\text{Ni}^\text{II}\text{Cu}^\text{II}$ for several values of the magnetic field. At zero magnetic field, the tripartite negativity gradually decreases upon increasing temperature when starting from a zero-temperature asymptotic limit \( N_\text{abc} = 5/8 \) and vanishing approximately at the threshold temperature $T \approx 37\,\text{K}$. The same general trends can be also detected for magnetic fields weaker than the first transition field \( B=22.8\,\text{T} \) (blue line marked with hexagons) favoring the singlet ground state \( |\psi_9\rangle \). { In the range of magnetic fields between $B=22.8\,\text{T}$ and $B=45.6\,\text{T}$} (black line marked with circles), the tripartite negativity starts from a lower local maximum of around \( N_\text{abc} \approx 0.454 \). The weaker value of the tripartite entanglement bears a close relation with presence of the triplet ground state \( |\psi_5\rangle \), but the respective threshold temperature remains unaffected by this change. A notable non-monotonous thermal dependence of the tripartite negativity is finally observed when the magnetic field slightly exceeds the saturation value \( B \approx 45.6\,\text{T} \). Under such conditions, the tripartite negativity \( N_{\text{abc}} \) exhibits a reentrant behavior when it first becomes nonzero at a lower threshold temperature, then rises steadily to a relatively low local maximum, which is successively followed by a gradual monotonous decline until it completely disappears at a higher threshold temperature (see the particular case for \( B = 50 \) T).

Figure \ref{fig:exp_Trip_Neg_Coh}(b) displays typical temperature dependencies of the tripartite quantum coherence \( C_\text{abc} \) of the mixed spin-(1/2,1,1/2) Heisenberg trimer by adapting the same set of parameters as used in Fig. \ref{fig:exp_Trip_Neg_Coh}(a) for a theoretical modeling of the tripartite entanglement of the molecular complex $\text{Cu}^\text{II}\text{Ni}^\text{II}\text{Cu}^\text{II}$. At low temperatures, the coherence  \( C_\text{abc} \) exhibits a quite analogous behavior to the tripartite negativity including three different zero-temperature asymptotic values assigned to the singlet ground state \( |\psi_9\rangle \), the triplet ground state \( |\psi_5\rangle \), and the fully polarized ground state \( |\psi_{11}\rangle \), respectively. In contrast to the tripartite negativity, the coherence \( C_\text{abc} \)  does not completely vanish at the threshold temperature but it persists of being nonzero in a high-temperature regime due to a more gradual thermally-induced decay. This observation bears evidence that the molecular nanomagnet $\text{Cu}^\text{II}\text{Ni}^\text{II}\text{Cu}^\text{II}$ retains a certain degree of quantum coherence even at elevated temperatures, which are well above the threshold temperature where the tripartite quantum entanglement is completely lost.

{ Next,  Fig. \ref{fig:exp_Bip_Neg} illustrates the temperature dependence of the bipartite negativity for the mixed spin-(1/2,1,1/2) Heisenberg trimer, which has been adapted to theoretically model the molecular complex $\text{Cu}^\text{II}\text{Ni}^\text{II}\text{Cu}^\text{II}$ under selected magnetic field strengths. The results presented in Fig. \ref{fig:exp_Bip_Neg}(a) for the bipartite negativity \( N_\text{ab} \) reveal that the bipartite entanglement between $\text{Cu}^\text{II}$ and $\text{Ni}^\text{II}$ magnetic ions remains sizable for temperatures \( T \lesssim 34\,\text{K} \) and exhibits trends  quite similar to those of the tripartite negativity \( N_\text{abc} \). This indicates that $\text{Cu}^\text{II}$ and $\text{Ni}^\text{II}$ magnetic ions are within the molecular compound $\text{Cu}^\text{II}\text{Ni}^\text{II}\text{Cu}^\text{II}$ subject to a strong bipartite entanglement at sufficiently low temperatures and magnetic fields promoting the singlet and triplet ground states \( |\psi_9\rangle \) and \( |\psi_5\rangle \), respectively. Indeed, the bipartite negativity \( N_\text{ab} \) gradually diminishes with increasing temperature until it completely vanishes at a threshold temperature of \( T \approx 34\,\text{K} \), which is comparable to the threshold temperature identified for the tripartite negativity \( N_\text{abc} \).

\begin{figure}[hbt!]
\centering
\includegraphics[scale=0.33]{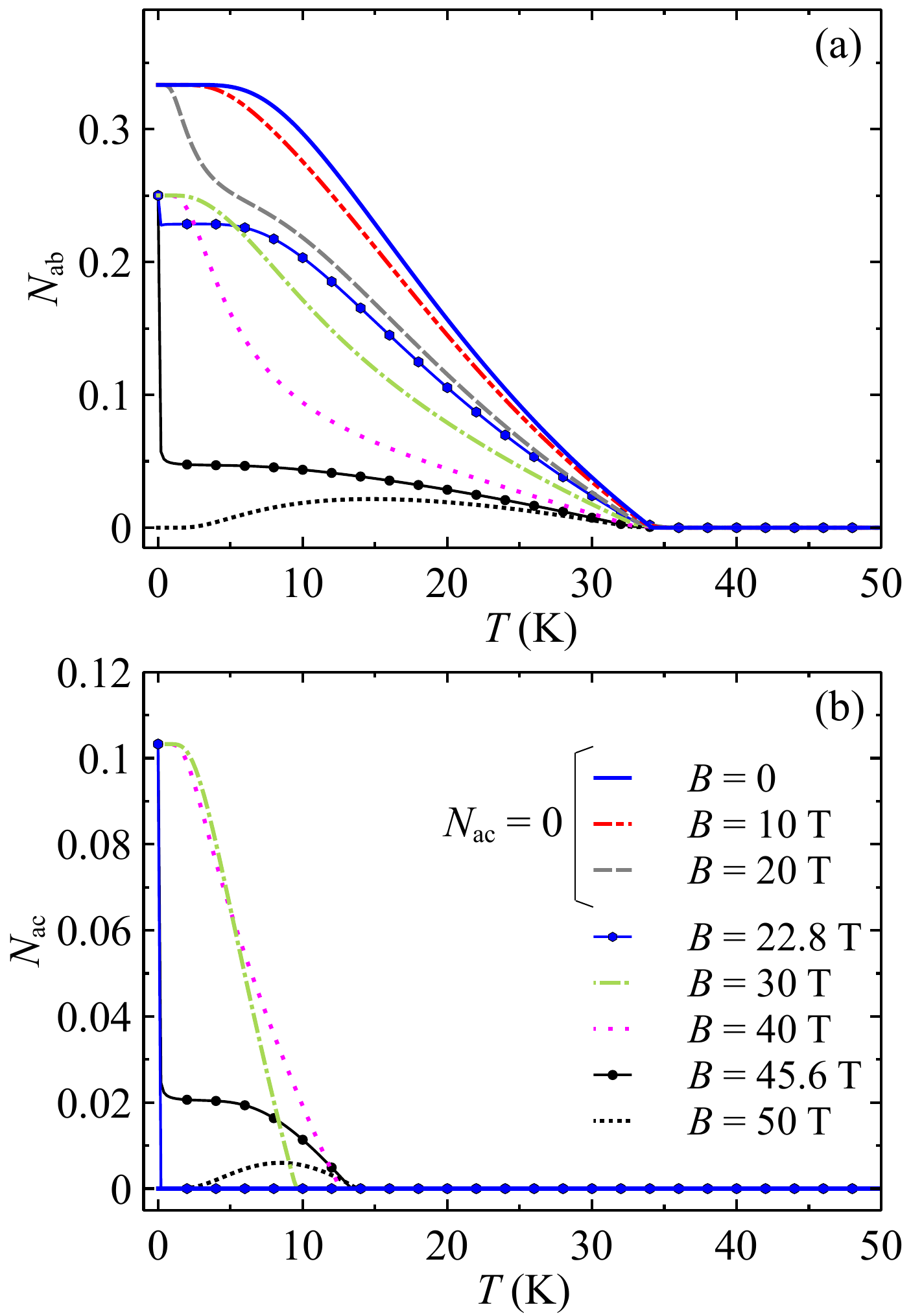}
\caption{{  Temperature dependencies of the bipartite negativities \( N_\text{ab} \) (a) and \( N_\text{ac} \) (b) of the mixed spin-(1/2,1,1/2) Heisenberg trimer with the coupling constant \( J = 22.8\,\text{cm}^{-1} \), the single-ion anisotropy \( D = 0.05\,\text{cm}^{-1} \), and the gyromagnetic ratio \( g = 2.227 \) adjusted according to Ref. \cite{Biswas2010} to a theoretical modeling of the molecular nanomagnet $\text{Cu}^\text{II}\text{Ni}^\text{II}\text{Cu}^\text{II}$.}}
\label{fig:exp_Bip_Neg}
\end{figure}

Furthermore, the bipartite negativity ${N}_{\text{ac}}$ of the mixed spin-(1/2,1,1/2) Heisenberg trimer adapted for a theoretical modeling of the molecular nanomagnet $\text{Cu}^\text{II}\text{Ni}^\text{II}\text{Cu}^\text{II}$ is depicted in Fig. \ref{fig:exp_Bip_Neg}(b) as a function of temperature for the same magnetic-field values. This measure of the bipartite entanglement between two outer $\text{Cu}^\text{II}$ magnetic ions is zero at sufficiently low magnetic fields ($B \lesssim 22.8\,\text{T}$), but becomes nonzero at low enough temperatures within the specific range of moderately strong magnetic field $22.8\,\text{T} \lesssim B \lesssim 45.6\,\text{T}$. This observation signals the absence of bipartite entanglement between the two outer $\text{Cu}^\text{II}$ magnetic ions in the singlet ground state \( |\psi_9\rangle \) and its emergence in the triplet ground state \( |\psi_5\rangle \). { The absence of bipartite entanglement between the two outer $\text{Cu}^\text{II}$ magnetic ions in the singlet ground state \( |\psi_9\rangle \) relates to the relatively weak easy-plane single-ion anisotropy $D =0.05\,\text{cm}^{-1}$, which would need to overcome the coupling constant \( J = 22.8\,\text{cm}^{-1} \) to induce at least a partial bipartite entanglement between the two outer $\text{Cu}^\text{II}$ magnetic ions. With increasing temperature, the bipartite negativity  ${N}_{\text{ac}}$ gradually diminishes and ultimately vanishes at a threshold temperature $T \approx 14\,\text{K}$ even when a moderately strong magnetic field drives the molecular nanomagnet $\text{Cu}^\text{II}\text{Ni}^\text{II}\text{Cu}^\text{II}$ to the triplet ground state \( |\psi_5\rangle \).}

Bearing all this in mind, we may proceed to a systematic classification of the quantum entanglement observed in the heterotrinuclear molecular complex $\text{Cu}^\text{II}\text{Ni}^\text{II}\text{Cu}^\text{II}$ following the categorization proposed in Ref. \cite{Sabin2008}. { Based on the results depicted in Fig. \ref{fig:exp_Trip_Neg_Coh}(a) and Fig. \ref{fig:exp_Bip_Neg} considering specific values of the coupling constant \( J = 22.8\,\text{cm}^{-1} \) and the uniaxial single-ion anisotropy $D = 0.05\,\text{cm}^{-1}$}, two distinct types of the quantum entanglement are identified in the molecular nanomagnet $\text{Cu}^\text{II}\text{Ni}^\text{II}\text{Cu}^\text{II}$. 
\begin{itemize}
\item[(I)] \textbf{Star-shaped singlet state \( |\psi_9\rangle \) (subtype 2-2):}  
At sufficiently low magnetic fields $B \lesssim 22.8\,\text{T}$, the nonzero tripartite negativity $N_\text{abc}$ quantifying entanglement among all three constituent $\text{Cu}^\text{II}\text{Ni}^\text{II}\text{Cu}^\text{II}$ magnetic ions is accompanied with nonzero bipartite negativity $N_\text{ab} = N_\text{bc}$ between $\text{Cu}^\text{II}$ and $\text{Ni}^\text{II}$ magnetic ions, whereas the bipartite negativity $N_\text{ac}$ between the two outer $\text{Cu}^\text{II}$ magnetic ions is zero. The corresponding singlet ground state \( |\psi_9\rangle \) thus belongs to the category of \textit{star-shaped} states \cite{Sabin2008,Plesch67,Plesch68}.
\item[(II)] \textbf{W-like triplet state \( |\psi_5\rangle \) (subtype 2-3):}  
At moderate values of the magnetic field $22.8\,\text{T} \lesssim B \lesssim 45.6\,\text{T}$, all entanglement measures including the tripartite negativity $N_\text{abc}$ and the bipartite negativities
$N_\text{ab}$, $N_\text{bc}$, and $N_\text{ac}$ become nonzero indicating the quantum entanglement between all three constituent $\text{Cu}^\text{II}\text{Ni}^\text{II}\text{Cu}^\text{II}$ magnetic ions. The corresponding triplet ground state \( |\psi_5\rangle \) can be accordingly categorized to \textit{W-like} states \cite{Sabin2008}.
\end{itemize}
In addition, it is worthwhile to remark that the thermal state emergent in a rather narrow range of temperatures $34\,\text{K} \lesssim T \lesssim 37\,\text{K}$, where the tripartite negativity is  $N_\text{abc}$ is nonzero even though all bipartite negativities $N_\text{ab}$, $N_\text{bc}$, and $N_\text{ac}$ vanish, has signatures \textit{GHZ-like} states (subtype 2-0) \cite{Sabin2008}.
}

\section{Conclusions}\label{sec:conclusions}

In the present paper, we have comprehensively investigated basic quantum characteristics of the mixed spin-(1/2,1,1/2) Heisenberg trimer in the presence of the external magnetic field, which was designed for theoretical modeling of molecular nanomagnets such as the heterotrinuclear coordination compound \(\text{Cu}^\text{II}\text{Ni}^\text{II}\text{Cu}^\text{II}\). By employing rigorous analytical and numerical calculations, we have examined distributions of bipartite and tripartite entanglement, and the \(l_1\)-norm of quantum coherence under the influence of an external magnetic field.  Although the density matrix elements determining these entanglement measures cannot be directly measured, it has been verified in Ref. \cite{str2020} that they can be connected to local observables such as the local magnetizations  \( \langle \hat{S}_z \rangle \), \( \langle \hat{\mu}_z \rangle \), and the local pair correlation functions such as \( \langle \hat{S}_z \hat{\mu}_z \rangle \), etc. These observables can be indirectly determined via experimental techniques such as inelastic neutron scattering, or correlated with measurable thermodynamic quantities like susceptibility and specific heat. This provides a pathway for experimental verification of our theoretical predictions.

The key finding stemming from the present study is that the mixed spin-(1/2,1,1/2) Heisenberg trimer may display significant bipartite and tripartite entanglement up to relatively high temperatures and magnetic fields, whereby the quantum coherence is maintained even at elevated temperatures compared to entanglement. This indicates that the investigated quantum spin system may retain a high degree of quantum coherence despite strong thermal fluctuations. The quantum features of the mixed spin-(1/2,1,1/2) Heisenberg trimer were connected to the singlet and triplet ground states, which emerge in a range of low and moderate magnetic fields, respectively. 

The mixed spin-(1/2,1,1/2) Heisenberg trimer was also adapted for a theoretical modeling of multipartite entanglement in the molecular nanomagnet \(\text{Cu}^\text{II}\text{Ni}^\text{II}\text{Cu}^\text{II}\). It has been verified that the molecular complex \(\text{Cu}^\text{II}\text{Ni}^\text{II}\text{Cu}^\text{II}\) exhibits a significant bipartite entanglement between $\text{Cu}^\text{II}$-$\text{Ni}^\text{II}$ magnetic ions and a significant tripartite entanglement among all three constituent $\text{Cu}^\text{II}$-$\text{Ni}^\text{II}$-$\text{Cu}^\text{II}$ magnetic ions, which persist up to relatively high temperatures $37\,\text{K}$ and magnetic fields $46\,\text{T}$. { Contrary to this, the bipartite entanglement between the two outer $\text{Cu}^\text{II}$ magnetic ions exist only in a range of moderate magnetic fields $23\,\text{T} \lesssim B \lesssim 46\,\text{T}$ and vanishes above much lower threshold temperature $T \approx 14$~K. The molecular compound \(\text{Cu}^\text{II}\text{Ni}^\text{II}\text{Cu}^\text{II}\) thus provides a fascinating quantum resource for the star-shaped state emergent within the singlet state at low magnetic fields $B \lesssim 23\,\text{T}$ and the W-like state found within the triplet state in a range of moderately strong magnetic fields $23\,\text{T} \lesssim B \lesssim 46\,\text{T}$.} In addition, it has been theoretically predicted that molecular nanomagnet \(\text{Cu}^\text{II}\text{Ni}^\text{II}\text{Cu}^\text{II}\) displays a high degree of the quantum coherence persisting even at elevated temperatures in spite of strong thermal fluctuations. 

This study particularly focused on the effects of easy-plane single-ion anisotropy and a longitudinal magnetic field on the quantum properties of the molecular compound $\text{Cu}^\text{II}\text{Ni}^\text{II}\text{Cu}^\text{II}$. However, exploring the impact of easy-axis anisotropy (\(D < 0\)) and applying the magnetic field in other directions could offer deeper insights into the behavior of quantum correlations in systems analogous to $\text{Cu}^\text{II}\text{Ni}^\text{II}\text{Cu}^\text{II}$. In future work, we plan to address these intriguing aspects which could significantly enhance our understanding of quantum features in small magnetic clusters.

\section*{Acknowledgements}
A.G. acknowledges Saeed’s Quantum Information Group (SSQIG) for accommodating hospitality during the course of this project. H.A.Z. acknowledges the financial support provided under the postdoctoral fellowship program of P. J. \v{S}af\'arik University in Ko\v{s}ice, Slovakia. S.H. was supported by Semnan University under Contract No. 21270. J.S. acknowledges financial support by the grant of Slovak Research and Development Agency provided under the contract Nos.  APVV-18-0197, APVV-22-0172 and by the grant of The Ministry of Education, Research, Development and Youth of the Slovak Republic under the contract No. VEGA 1/0695/23. V.O and Zh. A. acknowledge partial financial support form ANSEF (Grant No. PS-condmatth-2884) and from CS RA MESCS (Grants No. 21AG-1C047, 21AG-1C006 and 23AA-1C032).



\appendix

\begin{widetext}

\section{Eigenvectors and eigenenergies}
\label{Sec:appendix a}

Eigenvectors of the mixed spin-(1/2,1,1/2) Heisenberg trimer given by the Hamiltonian \ref{Eq:hamiltonian} are defined in Table \ref{tab:table1}, whereby the respective probability amplitudes entering into the relevant eigenvectors are specifically given in Table \ref{tab:table2}.
\begin{table}
\caption{\label{tab:table1} Eigenvectors, total quantum spin number $S_\text{T}^z$, and eigenenergies of the mixed spin-(1/2,1,1/2) Heisenberg trimer.}
\vspace{0.2cm}
\centering
\begin{tabular}{| c | c | c |}
\hline
Eigenvectors & $S_\text{T}^z$ & Eigenenergies \\ 
\hline
$\rvert \psi_{1} \rangle$ = $\frac{1}{\sqrt{2}} ( \rvert \downarrow 0 \uparrow \rangle - \rvert \uparrow 0 \downarrow \rangle$)  & 0 & $E_{1}=0$  \\
      \hline
 $\rvert \psi_{2} \rangle$ =  $\frac{1}{\sqrt{2}} ( \rvert \downarrow 1 \downarrow \rangle - \rvert \uparrow -1 \uparrow \rangle$)  & 0 & $E_2= D - J$ \\ 
      \hline
$\rvert \psi_3\rangle$ =  $\frac{1}{\sqrt{2}} ( \rvert \downarrow 1 \uparrow \rangle - \rvert \uparrow 1 \downarrow \rangle$)  & 1 & $E_3 = D - h $  \\ 
\hline
$\rvert \psi_4\rangle = \frac{1}{\sqrt{2}} ( \rvert \downarrow -1 \uparrow \rangle - \rvert \uparrow -1 \downarrow \rangle$) & -1 & $ E_4= D + h $ 
   \\  
\hline
$\rvert \psi_5\rangle = a (\rvert \uparrow 1 \downarrow \rangle + \rvert \downarrow 1 \uparrow \rangle) + b \rvert \uparrow 0 \uparrow \rangle $  & 1 & $E_5= \frac{D}{2} - h - \frac{1}{2}\sqrt{D^2 + 4 J^2} $ \\ 
\hline
$\rvert \psi_6\rangle = c (\rvert \uparrow 1 \downarrow \rangle + \rvert \downarrow 1 \uparrow \rangle) + d \rvert \uparrow 0 \uparrow \rangle $  & 1 & $E_6= \frac{D}{2} - h + \frac{1}{2}\sqrt{D^2 + 4 J^2} $ \\ 
\hline
$\rvert \psi_7\rangle = a (\rvert \uparrow -1 \downarrow \rangle + \rvert \downarrow -1 \uparrow \rangle) + b \rvert \downarrow 0 \downarrow \rangle $  & -1 & $E_7= \frac{D}{2} + h - \frac{1}{2}\sqrt{D^2 + 4 J^2} $ \\ 
\hline
$\rvert \psi_8\rangle = c (\rvert \uparrow -1 \downarrow \rangle + \rvert \downarrow -1 \uparrow \rangle) + d \rvert \downarrow 0 \downarrow \rangle $  & -1 & $E_8= \frac{D}{2} + h + \frac{1}{2}\sqrt{D^2 + 4 J^2} $ \\ 
\hline
$\rvert \psi_9\rangle = e (\rvert \uparrow 0 \downarrow \rangle + \rvert \downarrow 0 \uparrow \rangle) + f ( \rvert \uparrow -1 \uparrow \rangle + \rvert \downarrow 1 \downarrow \rangle )$  & 0 & $E_9= \frac{D-J}{2}- \sqrt{(\frac{D-J}{2})^2 + 2J^2}  $ \\ 
\hline
$\rvert \psi_{10}\rangle = g (\rvert \uparrow 0 \downarrow \rangle + \rvert \downarrow 0 \uparrow \rangle) + h ( \rvert \uparrow -1 \uparrow \rangle + \rvert \downarrow 1 \downarrow \rangle )$  & 0 & $E_{10} = \frac{D-J}{2}+\sqrt{(\frac{D-J}{2})^2 + 2J^2}  $ \\ 
\hline
$\rvert \psi_{11}\rangle = \rvert \uparrow 1 \uparrow \rangle $ &  2 & $E_{11} = D+J-2h  $  \\
\hline
$\rvert \psi_{12}\rangle = \rvert \downarrow -1 \downarrow \rangle $  &  -2 & $E_{12} = D+J+2h  $  \\
\hline
\end{tabular}
\end{table}

\begin{table}
\caption{\label{tab:table2} Probability amplitudes of the eigenvectors of the mixed spin-(1/2,1,1/2) Heisenberg trimer.}
\vspace{0.2cm}
\centering
\begin{tabular}{| c | c |}
\hline
 $a = \frac{\frac{D}{2} - \sqrt{J^2 + \left( \frac{D}{2}  \right)^2}}{\sqrt{2J^2 + 2 \left( \frac{D}{2} - \sqrt{J^2 + \left( \frac{D}{2}  \right)^2} \right)^2}} $ &
    $  b = \frac{J}{\sqrt{J^2 +  \left( \frac{D}{2} - \sqrt{ J^2 +(\frac{D}{2}) ^2}  \right)^2 }} $ 
 \\ 
      \hline
  $  c = \frac{\frac{D}{2} + \sqrt{J^2 + \left(\frac{D}{2}  \right)^2} }{\sqrt{2J^2 + 2 \left( \frac{D}{2} + \sqrt{J^2 + \left( \frac{D}{2}  \right)^2} \right)^2}}   $ &
$  d = \frac{J}{\sqrt{J^2 +  \left( \frac{D}{2} + \sqrt{ J^2 +(\frac{D}{2}) ^2}  \right)^2 }} $
  \\  
\hline
$ e = \frac{J}{ \sqrt{2J^2 +  \left( \frac{D-J}{2}  - \sqrt{ 2J^2 +(\frac{D - J}{2}) ^2}  \right)^2 }}  $ &
$ f = \frac{  \frac{D-J}{2} - \sqrt{ 2J^2 + \left( \frac{D-J}{2} \right)^2 }  }{\sqrt{4J^2 + 2 \left( \frac{D-J}{2} - \sqrt{ 2J^2 +(\frac{D - J}{2}) ^2}  \right)^2 }}$
  \\  
\hline
$ g = \frac{J}{ \sqrt{2J^2 +  \left( \frac{D-J}{2} + \sqrt{ 2J^2 +(\frac{D - J}{2}) ^2}  \right)^2 }}  $ &
$ h = \frac{  \frac{D-J}{2} + \sqrt{ 2J^2 + \left( \frac{D-J}{2} \right)^2 }  }{\sqrt{4J^2 + 2 \left( \frac{D-J}{2} + \sqrt{ 2J^2 +(\frac{D - J}{2}) ^2}  \right)^2 }}$
 \\ 
\hline
\end{tabular}
\end{table}

\section{The total, partially transposed and reduced density matrices} \label{Sec:appendix b}
The total density matrix $\rho$ of the mixed spin-(1/2,1,1/2) Heisenberg trimer reads:

\begin{align}  
\rho =
\begin{pmatrix}
\rho_{1,1} & 0 & 0 & 0 & 0 & 0 & 0 & 0 & 0 & 0 & 0 & 0 \\
0 & \rho_{2,2} & \rho_{2,3} & 0 & 0 &0 & \rho_{2,7} & 0 & 0 & 0 & 0 & 0 \\
0 & \rho_{2,3} & \rho_{3,3} & 0 & 0 &0 & \rho_{3,7} & 0 & 0 & 0 & 0 & 0 \\
0 & 0 & 0 & \rho_{4,4} & \rho_{4,5} & 0 & 0 & \rho_{4,8} & \rho_{4,9} & 0 & 0 & 0 \\
0 & 0 & 0 & \rho_{4,5} & \rho_{5,5} & 0 & 0 & \rho_{5,8} & \rho_{5,9} & 0 &  0 & 0 \\
0 & 0 & 0 & 0 & 0 & \rho_{6,6} & 0 & 0 & 0 & \rho_{6,10} & \rho_{6,11} & 0 \\
0 & \rho_{2,7} & \rho_{3,7} & 0 & 0 &0 & \rho_{7,7} & 0 & 0 & 0 & 0 & 0 \\
0 & 0 & 0 & \rho_{4,8} & \rho_{5,8} & 0 & 0 & \rho_{8,8} & \rho_{8,9} & 0 & 0 & 0 \\
0 & 0 & 0 & \rho_{4,9} & \rho_{5,9} & 0 & 0 & \rho_{8,9} & \rho_{9,9} & 0 & 0 & 0 \\
0 & 0 & 0 & 0 & 0 & \rho_{6,10} & 0 & 0 & 0 & \rho_{10,10} & \rho_{10,11} & 0 \\
0 & 0 & 0 & 0 & 0 & \rho_{6,11} & 0 & 0 & 0 & \rho_{10,11} & \rho_{11,11} & 0 \\
0 &0&0&0&0&0&0&0&0&0&0&\rho_{12,12}
\end{pmatrix}.
\nonumber
\end{align}  

The partially transposed density matrices with respect to each spin $s_\text{a}$, $S_\text{b}$ and $s_\text{c}$ are denoted by $\rho^{\text{T}_{\text{a}}}$, $\rho^{\text{T}_{\text{b}}}$ and $\rho^{\text{T}_{\text{c}}}$, given by:

\begin{align}  
\rho^{\text{T}_\text{a}} =
\begin{pmatrix}
\rho_{1,1} & 0 & 0 & 0 & 0 & 0 & 0 & \rho_{2,7} & \rho_{3,7} & 0 & 0 & 0 \\
0 & \rho_{2,2} & \rho_{2,3} & 0 & 0 &0 & 0 & 0 & 0 & \rho_{4,8} & \rho_{5,8} & 0 \\
0 & \rho_{2,3} & \rho_{3,3} & 0 & 0 &0 & 0 & 0 & 0 & \rho_{4,9} & \rho_{5,9} & 0 \\
0 & 0 & 0 & \rho_{4,4} & \rho_{4,5} & 0 & 0 & 0 & 0 & 0 & 0 & \rho_{6,10} \\
0 & 0 & 0 & \rho_{4,5} & \rho_{5,5} & 0 & 0 & 0 & 0 & 0 &  0 & \rho_{6,11} \\
0 & 0 & 0 & 0 & 0 & \rho_{6,6} & 0 & 0 & 0 & 0 & 0 & 0 \\
0 & 0 & 0 & 0 & 0 & 0 & \rho_{7,7} & 0 & 0 & 0 & 0 & 0 \\
\rho_{2,7} & 0 & 0 & 0 & 0 & 0 & 0 & \rho_{8,8} & \rho_{8,9} & 0 & 0 & 0 \\
\rho_{3,7} & 0 & 0 & 0 & 0 & 0 & 0 & \rho_{8,9} & \rho_{9,9} & 0 & 0 & 0 \\
0 & \rho_{4,8} & \rho_{4,9} & 0 & 0 & 0 & 0 & 0 & 0 & \rho_{10,10} & \rho_{10,11} & 0 \\
0 & \rho_{5,8} & \rho_{5,9} & 0 & 0 & 0 & 0 & 0 & 0 & \rho_{10,11} & \rho_{11,11} & 0 \\
0 &0&0&\rho_{6,10}&\rho_{6,11}&0&0&0&0&0&0&\rho_{12,12}
\end{pmatrix},
\nonumber
\end{align} 

\begin{eqnarray}  
\rho^{\text{T}_\text{b}} =
\begin{pmatrix}
\rho_{1,1} & 0 & 0 & \rho_{2,3} & 0 & 0 & 0 & 0 & \rho_{3,7} & 0 & 0 & \rho_{5,8} \\
0 & \rho_{2,2} & 0 & 0 & 0 &0 & \rho_{2,7} & 0 & 0 & \rho_{4,8} & 0 & 0 \\
0 & 0 & \rho_{3,3} & 0 & 0 & \rho_{4,5} & 0 & 0 & 0 & 0 & \rho_{5,9} & 0 \\
\rho_{2,3} & 0 & 0 & \rho_{4,4} & 0 & 0 & 0 & 0 & \rho_{4,9} & 0 & 0 & \rho_{6,10} \\
0 & 0 & 0 & 0 & \rho_{5,5} & 0 & 0 & 0 & 0 & 0 &  0 & 0 \\
0 & 0 & \rho_{4,5} & 0 & 0 & \rho_{6,6} & 0 & 0 & 0 & 0 & \rho_{6,11} & 0 \\
0 & \rho_{2,7} & 0 & 0 & 0 &0 & \rho_{7,7} & 0 & 0 & \rho_{8,9} & 0 & 0 \\
0 & 0 & 0 & 0 & 0 & 0 & 0 & \rho_{8,8} & 0 & 0 & 0 & 0 \\
\rho_{3,7} & 0 & 0 & \rho_{4,9} & 0 & 0 & 0 & 0 & \rho_{9,9} & 0 & 0 & \rho_{10,11} \\
0 & \rho_{4,8} & 0 & 0 & 0 & 0 & \rho_{8,9} & 0 & 0 & \rho_{10,10} & 0 & 0 \\
0 & 0 & \rho_{5,9} & 0 & 0 & \rho_{6,11} & 0 & 0 & 0 & 0 & \rho_{11,11} & 0 \\
\rho_{5,8} &0&0&\rho_{6,10}&0&0&0&0&\rho_{10,11}&0&0&\rho_{12,12}
\end{pmatrix},
\nonumber
\end{eqnarray}

\begin{eqnarray}  
\rho^{\text{T}_\text{c}} =
\begin{pmatrix}
\rho_{1,1} & 0 & 0 & \rho_{2,3} & 0 & 0 & 0 & \rho_{2,7} & 0 & 0 & 0 & 0 \\
0 & \rho_{2,2} & 0 & 0 & 0 &0 & 0 & 0 & 0 & 0 & 0 & 0 \\
0 & 0 & \rho_{3,3} & 0 & 0 & \rho_{4,5} & \rho_{37} & 0 & 0 & \rho_{4,9} & 0 & 0 \\
\rho_{2,3} & 0 & 0 & \rho_{4,4} & 0 & 0 & 0 & \rho_{4,8} & 0 & 0 & 0 & 0 \\
0 & 0 & 0 & 0 & \rho_{5,5} & 0 & 0 & 0 & \rho_{5,9} & 0 &  0 & \rho_{6,11} \\
0 & 0 & \rho_{4,5} & 0 & 0 & \rho_{6,6} & \rho_{5,8} & 0 & 0 & \rho_{6,10} & 0 & 0 \\
0 & 0 & \rho_{3,7} & 0 & 0 & \rho_{5,8} & \rho_{7,7} & 0 & 0 & \rho_{8,9} & 0 & 0 \\
\rho_{2,7} & 0 & 0 & \rho_{4,8} & 0 & 0 & 0 & \rho_{8,8} & 0 & 0 & 0 & 0 \\
0 & 0 & 0 & 0 & \rho_{5,9} & 0 & 0 & 0 & \rho_{9,9} & 0 & 0 & \rho_{10,11} \\
0 & 0 & \rho_{4,9} & 0 & 0 & \rho_{6,10} & \rho_{8,9} & 0 & 0 & \rho_{10,10} & 0 & 0 \\
0 & 0 & 0 & 0 & 0 & 0 & 0 & 0 & 0 & 0 & \rho_{11,11} & 0 \\0 &0&0&0&\rho_{6,11}&0&0&0&\rho_{10,11}&0&0&\rho_{12,12}
\end{pmatrix}.
\nonumber
\end{eqnarray} 

The reduced density matrix $\varrho_{\text{ab}}$ is given by:

\begin{eqnarray}  
\varrho_{\text{ab}} \!=\!\! 
\begin{pmatrix}
\rho_{1,1}+\rho_{2,2}&0&0&0&0&0\\
 0 & \rho_{3,3}+\rho_{4,4} & 0 & \rho_{2,3} + \rho_{4,5} & 0 & 0 \\
 0 & 0 & \rho_{5,5}+\rho_{6,6} & 0 & \rho_{4,5} + \rho_{6,10} & 0 \\
 0 & \rho_{2,3} + \rho_{4,5} & 0 & \rho_{7,7}+\rho_{8,8} & 0 & 0  \\
0 & 0 & \rho_{4,5} + \rho_{6,10} & 0 & \rho_{9,9}+\rho_{10,10} & 0 \\ 0 & 0 & 0 & 0 & 0 & \rho_{11,11}+\rho_{12,12} 
\end{pmatrix}.
\nonumber
\end{eqnarray}  
The partially transposed reduced density matrix $\varrho_{\text{ab}}^{\text{T}_{\text{b}}}$ is as follows:

\begin{eqnarray}  
\varrho_{\text{ab}}^{\text{T}_\text{b}} = 
\begin{pmatrix}
\rho_{1,1}+\rho_{2,2}&0&0&0&\rho_{2,3} + \rho_{4,5}& 0 \\
 0 & \rho_{3,3}+\rho_{4,4} & 0 & 0 & 0 & \rho_{4,5} + \rho_{6,10} \\
 0 & 0 & \rho_{5,5}+\rho_{6,6} & 0 & 0 & 0 \\
 0 & 0 & 0 & \rho_{7,7}+\rho_{8,8} & 0 & 0  \\
\rho_{2,3} + \rho_{4,5} & 0 & 0 & 0 & \rho_{9,9}+\rho_{10,10} & 0 \\ 0 & 
\rho_{4,5} + \rho_{6,10} & 0 & 0 & 0 & \rho_{11,11}+\rho_{12,12} 
\end{pmatrix}.
\nonumber
\end{eqnarray}  

The reduced density matrix $\varrho_{\text{bc}}$ is given by:

\begin{eqnarray}  
\varrho_{\text{bc}} \!=\!\! 
\begin{pmatrix}
\rho_{1,1}+\rho_{7,7}&0&0&0&0&0\\
 0 & \rho_{2,2}+\rho_{8,8} & \rho_{2,3} + \rho_{4,5} & 0 & 0 & 0 \\
 0 & \rho_{2,3} + \rho_{8,9} & \rho_{3,3}+\rho_{9,9} & 0 & 0 & 0 \\
 0 & 0 & 0 & \rho_{4,4}+\rho_{10,10} & \rho_{4,5} + \rho_{10,11} & 0  \\
0 & 0 & 0 & \rho_{4,5} + \rho_{10,11} & \rho_{5,5}+\rho_{11,11} & 0 \\ 0 & 0 & 0 & 0 & 0 & \rho_{6,6}+\rho_{12,12} 
\end{pmatrix}.
\nonumber
\end{eqnarray}  

The partially transposed reduced density matrix $\varrho_{\text{bc}}^{\text{T}_{\text{b}}}$ is as follows:

\begin{eqnarray}  
\varrho_{\text{bc}}^{\text{T}_\text{b}} = 
\begin{pmatrix}
\rho_{1,1}+\rho_{7,7}&0&0&\rho_{2,3} + \rho_{8,9}&0& 0 \\
 0 & \rho_{2,2}+\rho_{8,8} & 0 & 0 & 0 & 0 \\
 0 & 0 & \rho_{3,3}+\rho_{9,9} & 0 & 0 & \rho_{4,5} + \rho_{10,11} \\
 \rho_{2,3} + \rho_{8,9} & 0 & 0 & \rho_{4,4} + \rho_{10,10} & 0 & 0  \\
0 & 0 & 0 & 0 & \rho_{5,5}+\rho_{11,11} & 0 \\ 0 & 
0 & \rho_{4,5} + \rho_{10,11} & 0 & 0 & \rho_{6,6}+\rho_{12,12} 
\end{pmatrix}.
\nonumber
\end{eqnarray}  

The reduced density matrix $\varrho_{\text{ac}}$ would be given by the following formula:
\begin{eqnarray}  
\varrho_{\text{ac}} =
\begin{pmatrix}
\rho_{1,1}+\rho_{3,3} + \rho_{5,5}& 0 & 0 & 0  \\
 0 & \rho_{2,2}+\rho_{4,4} + \rho_{6,6} & \rho_{2,7} + \rho_{4,9} + \rho_{6,11} & 0 \\
 0 & \rho_{2,7} + \rho_{4,9} + \rho_{6,11} & \rho_{7,7}+\rho_{9,9}+\rho_{11,11} & 0 \\
 0 & 0 & 0 & \rho_{8,8}+ \rho_{10,10} + \rho_{12,12} 
\end{pmatrix}.
\nonumber
\end{eqnarray}  

The partially transposed density matrix $\varrho_{\text{ac}}^{\text{T}_\text{c}}$ takes the form:
\begin{eqnarray}  
\varrho_{\text{ac}}^{\text{T}_\text{c}} =
\begin{pmatrix}
\rho_{1,1}+\rho_{3,3} + 
\rho_{5,5}& 0 & 0 &\rho_{2,7} + \rho_{4,9} + \rho_{6,11} \\
 0 & \rho_{2,2}+\rho_{4,4} + \rho_{6,6} & 0 & 0 \\
 0 & 0 & \rho_{7,7}+\rho_{9,9}+\rho_{11,11} & 0 \\
 \rho_{2,7} + \rho_{4,9} + \rho_{6,11} & 0 & 0 & \rho_{8,8}+ \rho_{10,10} + \rho_{12,12} 
\end{pmatrix}.
\nonumber
\end{eqnarray}

The elements of the density matrix are presented below where $Z$ is the partition function given in formula \eqref{Eq:part_func}:

\begin{eqnarray*}
\rho_{1,1} &=& \frac{1}{Z}  \exp\left(- \beta {E_{11}}\right), \\
\rho_{2,2} &=& \rho_{7,7} = \frac{1}{Z} \left [\frac{1}{2} \exp\left(- \beta E_{3}\right) + a^2 \exp\left(- \beta E_{5}\right) + c^2 \exp\left(- \beta E_{6} \right)  \right],\\
\rho_{2,3} &=&  \rho_{3,7} = \frac{1}{Z}\left[a b \exp\left(- \beta E_{5}\right) + c d \exp\left(- \beta E_{6}\right) \right], \\
\rho_{2,7} &=& \frac{1}{Z} \left[- \frac{1}{2} \exp\left(- \beta E_{3}\right) + a^2 \exp\left(- \beta E_{5}\right) + c^2 \exp\left(- \beta E_{6} \right) \right], \\
\rho_{3,3} &=&\frac{1}{Z}\left[ b^2 \exp\left(- \beta E_{5}\right) + d^2 \exp\left(- \beta E_{6} \right) \right],\\
\rho_{4,4} &=&  \rho_{99} =\frac{1}{Z}\left[ \frac{1}{2} \exp\left(- \beta E_{1}\right) + e^2 \exp\left(- \beta E_{9}\right) + g^2 \exp\left(- \beta E_{10}\right)\right], \\
\rho_{4,5} &=& \rho_{4,8} =  
\rho_{5,9} = \rho_{8,9} =  \frac{1}{Z} \left[ef\exp\left(- \beta E_{9}\right) + g h \exp\left(- \beta E_{10}\right)\right], \\
\rho_{4,9} &=& \frac{1}{Z} \left[- \frac{1}{2} \exp\left(- \beta E_{1}\right) + e^2 \exp\left(- \beta E_{9}\right) + g^2 \exp\left(- \beta E_{10}\right) \right],\\
\rho_{5,5} &=&\rho_{88} =\frac{1}{Z} \left[ \frac{1}{2} \exp\left(- \beta E_{2}\right) + f^2 \exp\left(- \beta E_{9}\right) + h^2 \exp\left(- \beta E_{10}\right)\right], \\ 
\rho_{5,8} &=& \frac{1}{Z} \left[ - \frac{1}{2} \exp\left( -\beta E_{2} \right) + f^2 \exp\left(-\beta E_{9}\right) + h^2 \exp\left(-\beta E_{10}\right)   \right],
\\ 
\rho_{6,6} &=& \rho_{11,11} = \frac{1}{Z} \left[ \frac{1}{2} \exp\left(- \beta E_{4}\right) + a^2 \exp\left(- \beta E_{7}\right) + c^2 \exp\left(- \beta E_{8}\right) \right],\\
\rho_{6,10} &=& \rho_{10,11} =\frac{1}{Z} \left[ a b \exp\left(- \beta E_{7}\right) + c d \exp\left(- \beta E_{8}\right)\right], \\
\rho_{6,11} &=&\frac{1}{Z} \left[ - \frac{1}{2} \exp\left(- \beta E_{4}\right) + a^2 \exp\left(- \beta E_{7}\right) + c^2 \exp\left(- \beta E_{8}\right)\right], \\
\rho_{10,10} &=& \frac{1}{Z}\left[b^2 \exp\left(- \beta E_{7}\right) + d^2 \exp\left(- \beta E_{8}\right) \right],\\
\rho_{12,12} &=& \frac{1}{Z}\exp\left(- \beta E_{12}\right).
\end{eqnarray*}

\section{Eigenvalues of partially transposed reduced density matrices} \label{Sec:appendix c}

All eigenvalues of the partially transposed reduced density matrix $\varrho_{\text{ab}}^{\text{T}_{\text{b}}}$ that constitute bipartite negativity $N_{\text{ab}}$ (according to the formula) are as follows:
\begin{eqnarray*}
  \lambda_{1} =  \rho_{2,2} + \rho_{8,8}, \qquad \lambda_{2} =  \rho_{5,5} + \rho_{11,11},
\end{eqnarray*}

\begin{eqnarray*}
    \lambda_{3;4} = \frac{1}{2} \left( \rho_{1,1} + \rho_{7,7} + \rho_{4,4} + \rho_{10,10} \pm \sqrt{(\rho_{1,1} + \rho_{7,7} - \rho_{4,4} - \rho_{10,10})^2 + 4 (\rho_{2,3} + \rho_{8,9}) ^ 2}
 \right), 
 \end{eqnarray*}
and
\begin{eqnarray*}
  \lambda_{5;6} = \frac{1}{2} \left( \rho_{3,3} + \rho_{9,9} + \rho_{6,6} + \rho_{12,12} \pm \sqrt{(\rho_{3,3} + \rho_{9,9} - \rho_{6,6} - \rho_{12,12})^2 + 4 (\rho_{10,11} + \rho_{4,5}) ^ 2}
 \right).
\end{eqnarray*}

Besides, the eigenvalue of the partially transposed reduced density matrix $\varrho_{\text{ac}}^{\text{T}_{\text{c}}}$ that constitutes bipartite negativity $N_{\text{ac}}$ would be expressed by:

\begin{eqnarray*}
  \lambda_{1} =  \rho_{2,2} + \rho_{4,4}+\rho_{6,6}, \qquad \lambda_{2} =  \rho_{7,7} +\rho_{9,9}+ \rho_{11,11},
\end{eqnarray*}
and
\begin{eqnarray*}
    \lambda_{3;4} = \frac{1}{2} \left( \rho_{1,1} + \rho_{3,3} + \rho_{5,5} + \rho_{8,8} + \rho_{10,10} + \rho_{12,12} \pm \sqrt{(\rho_{1,1} + \rho_{3,3} + \rho_{5,5} - \rho_{8,8} - \rho_{10,10} - \rho_{12,12})^2 + 4 (\rho_{2,7} + \rho_{4,9} + \rho_{6,11}) ^ 2} \right).
\end{eqnarray*}
\end{widetext}


\end{document}